\documentclass[twoside,11pt]{article} 
\usepackage[T1]{fontenc}
\usepackage{jmlr2e}
\usepackage{url}

\title{MLPACK: A Scalable C++ Machine Learning Library}

\author{\name Ryan R.~Curtin \email ryan.curtin@cc.gatech.edu \\
\name James R.~Cline \email james.cline@gatech.edu \\
\name N.~P.~Slagle \email npslagle@gmail.com \\
\name William B.~March \email march@gatech.edu \\
\name Parikshit Ram \email p.ram@gatech.edu \\
\name Nishant A.~Mehta \email niche@cc.gatech.edu \\
\name Alexander G.~Gray \email agray@cc.gatech.edu \\
\addr College of Computing \\
Georgia Institute of Technology\\
Atlanta, GA 30332 \\
}

\editor{ }

\begin{document}

\jmlrheading{1}{2012}{1-4}{9/12}{x/12}{Ryan R.~Curtin, James R.~Cline,
N.~P.~Slagle, William B.~March, P.~Ram, Nishant A.~Mehta, and A.~G.~Gray}

\ShortHeadings{MLPACK: A Scalable C++ Machine Learning Library}{Curtin, Cline,
Slagle, March, Ram, Mehta, and Gray}

\firstpageno{1}

\maketitle

\vspace*{-1.5em}

\begin{abstract}%
MLPACK is a state-of-the-art, scalable, multi-platform C++ machine learning
library released in late 2011 offering both a simple, consistent API accessible
to novice users and high performance and flexibility to expert users by
leveraging modern features of C++.  MLPACK provides cutting-edge algorithms
whose benchmarks exhibit far better performance than other leading machine
learning libraries.  MLPACK version 1.0.3, licensed under the LGPL, is available
at \texttt{http://www.mlpack.org}.
\end{abstract}

\vspace*{-0.5em}

\section{Introduction and Goals}


Though several machine learning libraries are freely available online, few,
if any, offer efficient algorithms to the average user.  For instance, the
popular Weka toolkit \citep{weka} emphasizes ease of use but scales
poorly; the distributed Apache Mahout library offers scalability at a cost of
higher overhead (such as clusters and powerful servers often unavailable to the
average user). Also, few libraries offer breadth; for instance, libsvm
\citep{libsvm} and the Tilburg Memory-Based Learner (TiMBL) are highly scalable
and accessible yet each offer only a single method.

MLPACK, intended to be the machine learning analog to the general-purpose
LAPACK linear algebra library, aims to combine efficiency and
accessibility.  Written in C++, MLPACK uses the highly efficient Armadillo
matrix library \citep{armadillo} and is freely available under the GNU Lesser
General Public License (LGPL).  Through the use of C++ templates, MLPACK both
eliminates unnecessary copying of datasets and performs expression
optimizations unavailable in other languages.  Also, MLPACK is, to our
knowledge, unique among existing libraries in using generic programming
features of C++ to allow customization of the available machine learning
methods without incurring performance penalties.

In addition, users ranging from students to experts should find the consistent,
intuitive interface of MLPACK to be highly accessible.  Finally, the source code
provides references and comprehensive documentation.

Four major goals of the development team of MLPACK are

\vspace*{-0.4em}

\begin{itemize}

  \item to implement \textbf{scalable, fast} machine learning algorithms,
  \vspace*{-1em}
  \item to design an \textbf{intuitive, consistent, and simple} API for non-expert users,
  \vspace*{-1em}
  \item to implement a \textbf{wide variety} of machine learning methods, and
  \vspace*{-1em}
  \item to provide \textbf{cutting-edge} machine learning algorithms unavailable elsewhere.

\end{itemize}

\vspace*{-0.4em}

This paper offers both an introduction to the simple and extensible API and
a glimpse of the superior performance of the library.

\section{Package Overview}

Each algorithm available in MLPACK features both a set of C++ library
functions and a standalone command-line executable.  Version 1.0.3 includes the
following methods:

\vspace*{-0.4em}

\begin{itemize}
\item nearest/furthest neighbor search with cover trees or kd-trees ($k$-nearest-neighbors)
\vspace*{-1em}
\item range search with cover trees or kd-trees
\vspace*{-1em}
\item Gaussian mixture models (GMMs)
\vspace*{-1em}
\item hidden Markov models (HMMs)
\vspace*{-1em}
\item LARS / Lasso regression
\vspace*{-1em}
\item k-means clustering
\vspace*{-1em}
\item fast hierarchical clustering (Euclidean MST calculation) \emph{(*)} \citep{2010march}
\vspace*{-1em}
\item kernel PCA (and regular PCA)
\vspace*{-1em}
\item local coordinate coding \emph{(*)} \citep{2009lcc}
\vspace*{-1em}
\item sparse coding using dictionary learning
\vspace*{-1em}
\item RADICAL (Robust, Accurate, Direct ICA aLgorithm)
\vspace*{-1em}
\item maximum variance unfolding (MVU) via LRSDP \emph{(*)} \citep{2003burer}
\vspace*{-1em}
\item the naive Bayes classifier
\vspace*{-1em}
\item density estimation trees \emph{(*)} \citep{2011ram}
\end{itemize}

\vspace*{-0.4em}

\emph{(*)}: algorithm is not available in any other comparable software package
\\

\vspace*{-0.4em}

The development team manages MLPACK with Subversion and the Trac bug reporting
system, allowing easy downloads and simple bug reporting.  The entire
development process is transparent, so any interested user can easily contribute
to the library.  MLPACK can compile from source on Linux, Mac OS, and Windows;
currently, different Linux distributions are reviewing MLPACK for inclusion in
their package managers, which will allow users to install MLPACK without needing
to compile from source.

\section{A Consistent, Simple API}

MLPACK features a highly accessible API, both in style (such as consistent
naming schemes and coding conventions) and ease of use (such as templated
defaults), as well as stringent documentation standards.  Consequently, a new
user can execute algorithms out-of-the-box often with little or no adjustment to
parameters, while the seasoned expert can expect extreme flexibility in
algorithmic tuning.  For example, the following line initializes an object which
will perform the standard k-means clustering in Euclidean space: \\

\vspace*{-0.7em}
\noindent{\small \texttt{KMeans<> k();}} \\
\vspace*{-0.7em}

However, an expert user could easily use the Manhattan distance, a different
cluster initialization policy, and allow empty clusters: \\

\vspace*{-0.7em}
\noindent{\small \texttt{KMeans<ManhattanDistance, KMeansPlusPlusInitialization, AllowEmptyClusters>
k();}} \\
\vspace*{-0.7em}

Users can implement these custom classes in their code, then simply
link against the MLPACK library, requiring no modification within the MLPACK
library.  In addition to this flexibility, Armadillo 3.4.0 includes sparse
matrix support; sparse matrices can be used in place of dense matrices for the
appropriate MLPACK methods.


\begin{table}[t]
\begin{center}
\small{
\begin{tabular}{| l | r@{.}l | r@{.}l | r@{.}l | r@{.}l | r@{.}l | r@{.}l |}
\hline
Dataset                 & \multicolumn{2}{|c|}{MLPACK} & \multicolumn{2}{|c|}{Weka} & \multicolumn{2}{|c|}{Shogun} & \multicolumn{2}{|c|}{MATLAB} & \multicolumn{2}{|c|}{mlpy} & \multicolumn{2}{|c|}{sklearn} \\
\hline
wine                    &    \textbf{0}&\textbf{0003} &        0&0621             &         0&0277              & 0&0021                      & 0&0025                    & 0&0008 \\
cloud                   &    \textbf{0}&\textbf{0069} &        0&1174             &         0&5000              & 0&0210                      & 0&3520                    & 0&0192 \\
wine-qual               &    \textbf{0}&\textbf{0290} &        0&8868             &         4&3617              & 0&6465                      & 4&0431                    & 0&1668 \\
isolet                  &   \textbf{13}&\textbf{0197} &      213&4735             &        37&6190              & 46&9518                     & 52&0437                   & 46&8016 \\
miniboone               &   \textbf{20}&\textbf{2045} &      216&1469             &      2351&4637              & 1088&1127                   & 3219&2696                 & 714&2385 \\
yp-msd                  & \textbf{5430}&\textbf{0478} &    >9000&0000             &     >9000&0000              & >9000&0000                  & >9000&0000                & >9000&0000 \\
corel                   &    \textbf{4}&\textbf{9716} &       14&4264             &       555&9600              & 60&8496                     & 209&5056                  & 160&4597 \\
covtype                 &   \textbf{14}&\textbf{3449} &       45&9912             &     >9000&0000              & >9000&0000                  & >9000&0000                & 651&6259 \\
mnist                   & \textbf{2719}&\textbf{8087} &    >9000&0000             &      3536&4477              & 4838&6747                   & 5192&3586                 & 5363&9650 \\
randu                   & \textbf{1020}&\textbf{9142} &     2665&0921             &     >9000&0000              & 1679&2893                   & >9000&0000                & 8780&0176 \\
\hline
\end{tabular}
}
\end{center}
\vspace*{-1.5em}
\caption{$k$-NN benchmarks (in seconds).}
\label{tab:knn}
\end{table}

\begin{table}
\begin{center}
\small{
\begin{tabular}{| l | c | c | c | c | c |}
\hline
Dataset & \textbf{wine} & \textbf{cloud} & \textbf{wine-qual} & \textbf{isolet} & \textbf{miniboone} \\
\hline
UCI Name & Wine & Cloud & Wine Quality & ISOLET & MiniBooNE \\
\hline
Size & 178x13 & 2048x10 & 6497x11 & 7797x617 & 130064x50  \\
\hline
\multicolumn{6}{c}{ } \\
\hline
Dataset & \textbf{yp-msd} & \textbf{corel} & \textbf{covtype} & \textbf{mnist} & \textbf{randu} \\
\hline
UCI Name & YearPredictionMSD & Corel & Covertype & \textit{N/A} & \textit{N/A} \\
\hline
Size & 515345x90 & 37749x32 & 581082x54 & 70000x784 & 1000000x10 \\
\hline
\end{tabular}
}
\end{center}
\vspace*{-1.5em}
\caption{Benchmark dataset sizes.}
\label{tab:datasets}
\vspace*{-0.7em}
\end{table}

\begin{table}[t]
\vspace*{-2.5em}
\begin{center}
\small{
\begin{tabular}{| l | c | r@{.}l | r@{.}l | r@{.}l | r@{.}l |}
\hline
Dataset           & Clusters & \multicolumn{2}{|c|}{MLPACK} & \multicolumn{2}{|c|}{Shogun} & \multicolumn{2}{|c|}{MATLAB} & \multicolumn{2}{|c|}{sklearn} \\
\hline
wine              &     3    &     \textbf{0}&\textbf{0006} &            0&0073            &           0&0055             &         \ \ \ 0&0064             \\
cloud             &     5    &     \textbf{0}&\textbf{0036} &            0&1240            &           0&0194             &         \ \ \ 0&1753             \\
wine-qual         &     7    &     \textbf{0}&\textbf{0221} &            0&6030            &           0&0987             &         \ \ \ 4&0407             \\
isolet            &    26    &     \textbf{4}&\textbf{9762} &            8&5093            &          54&7463             &         \ \ \ 7&0902             \\
miniboone         &     2    &     \textbf{0}&\textbf{1853} &            8&0206            &           0&7221             & \multicolumn{2}{|c|}{\textit{memory}} \\
yp-msd            &    10    &    \textbf{34}&\textbf{8223} &          135&8853            &         269&7302             & \multicolumn{2}{|c|}{\textit{memory}} \\
corel             &    10    &     \textbf{0}&\textbf{4672} &            2&4237            &           1&6318             & \multicolumn{2}{|c|}{\textit{memory}} \\
covtype           &     7    &    \textbf{13}&\textbf{5997} &           71&1283            &          54&9034             & \multicolumn{2}{|c|}{\textit{memory}} \\
mnist             &    10    &    \textbf{80}&\textbf{2092} &          163&7513            &         133&9970             & \multicolumn{2}{|c|}{\textit{memory}} \\
randu             &    75    &   \textbf{727}&\textbf{1498} &         7443&2675            &        3117&5177             & \multicolumn{2}{|c|}{\textit{memory}} \\
\hline
\end{tabular}
}
\end{center}
\vspace*{-1.5em}
\caption{$k$-means benchmarks (in seconds).}
\label{tab:kmeans}
\vspace*{-1.5em}
\end{table}

\section{Benchmarks}

To demonstrate the efficiency of the algorithms implemented in MLPACK, we
present a comparison of the running times of $k$-nearest-neighbors and the
$k$-means clustering algorithm from MLPACK, Weka \citep{weka}, MATLAB, the
Shogun Toolkit \citep{shogun}, mlpy \citep{mlpy}, and scikit.learn (`sklearn')
\citep{sklearn}, using a modest consumer-grade workstation containing an AMD
Phenom II X6 1100T processor clocked at $3.3$ GHz and 8 GB of RAM.

Eight datasets from the UCI datasets repository \citep{uci} are used; the MNIST
handwritten digit database is also used (`mnist') \citep{mnist}, as well as a
uniformly distributed random dataset (`randu').  Information on the sizes of
these ten datasets appears in Table \ref{tab:datasets}.  Dataset loading time is
not included in the benchmarks.  Each test was run 5 times; the average is shown
in the results.

$k$-NN was run with each library on each dataset, with $k = 3$.  The results for
each library and each dataset appears in Table \ref{tab:knn}.  The $k$-means
algorithm was run with the same starting centroids for each library, and
$1000$ iterations maximum.  The number of clusters $k$ was chosen to reflect the
structure of the dataset.  Benchmarks for $k$-means are given in Table
\ref{tab:kmeans}.  Weka and mlpy are excluded because they do not allow
specification of the starting centroids.  `\textit{memory}' indicates that the
system ran out of memory during the test.

MLPACK's $k$-nearest neighbors and $k$-means are faster than the competitors in
all test cases.  Benchmarks for other methods, omitted due to space constraints,
also show similar speedups over competing implementations.

\section{Future Plans and Conclusion}


The favorable benchmarks exhibited above are not necessarily the global optimum;
MLPACK's active development team includes several core developers and many
contributors.  Because MLPACK is open-source, contributions from outsiders
are welcome, including feature requests and bug reports.  Thus, the performance,
extensibility, and breadth of algorithms within MLPACK are all certain to improve.

The first releases of MLPACK lacked parallelism, but experimental parallel code
using OpenMP is currently in testing.  This parallel support must maintain a
simple API and avoid large, reverse-incompatible API changes.  Other useful
planned features include using on-disk databases (rather than requiring loading
the dataset entirely into RAM) and validation of saved models (such as trees or
distributions).  Refactoring work continues on existing code, providing more
flexible abstractions and greater extensibility.  Nevertheless, MLPACK's future
growth will mostly be the addition of new machine learning methods; since the
original release (1.0.0), there are five new methods.

In conclusion, we have shown that MLPACK is a state-of-the-art C++ machine
learning library which leverages the powerful C++ concept of generic programming
to give excellent performance on large datasets.

\section*{Acknowledgements}

A full list of developers and researchers (other than the authors) who have
contributed significantly to MLPACK are Sterling Peet, Vlad Grantcharov, Ajinkya
Kale, Dongryeol Lee, Chip Mappus, Hua Ouyang, Long Quoc Tran, Noah Kauffman,
Rajendran Mohan, and Trironk Kiatkungwanglai.

{
\bibliography{mlpack}
}

\end{document}